# INTERPRETING UNCERTAINTY IN MODEL PREDICTIONS FOR COVID-19 DIAGNOSIS

*Gayathiri Murugamoorthy, Naimul Khan*

Electrical, Computer, and Biomedical Engineering, Ryerson University, Toronto, ON

**ABSTRACT**

COVID-19, due to its accelerated spread has brought in the need to use assistive tools for faster diagnosis in addition to typical lab swab testing. Chest X-Rays for COVID cases tend to show changes in the lungs such as ground glass opacities and peripheral consolidations which can be detected by deep neural networks. However, traditional convolutional networks use point estimate for predictions, lacking in capture of uncertainty, which makes them less reliable for adoption. There have been several works so far in predicting COVID positive cases with chest X-Rays. However, not much has been explored on quantifying the uncertainty of these predictions, interpreting uncertainty, and decomposing this to model or data uncertainty. To address these needs, we develop a visualization framework to address interpretability of uncertainty and its components, with uncertainty in predictions computed with a Bayesian Convolutional Neural Network. This framework aims to understand the contribution of individual features in the Chest-X-Ray images to predictive uncertainty. Providing this as an assistive tool can help the radiologist understand why the model came up with a prediction and whether the regions of interest captured by the model for the specific prediction are of significance in diagnosis. We demonstrate the usefulness of the tool in chest x-ray interpretation through several test cases from a benchmark dataset.

## 1. INTRODUCTION

Chest X-Ray images and CT scans for COVID cases contain useful information for early diagnosis such as ground glass opacities, infiltrate air space opacities, peripheral consolidations in majority of the cases, and pleural effusion to be a very rare worst-case scenario. Since there is a huge variation at the same time similar features that can attribute to a type of pneumonia or any other respiratory disease, it is important to provide radiologists means to visualize why the model made a decision in addition to just being a classifier detecting COVID. This could help medical professionals verify model correctness and apply necessary measures to the dataset or the model for improving performance.

There have been several research works using point estimate neural networks for classifying COVID cases. These works range from binary classification [1][2][3][6], 3-class classification-COVID/Normal/Viral[1][2][3][4][5] and 4 class classification-COVID/Normal/Viral/Bacterial[2][3][4]. These works used a range of existing state of the art models such as CheXNet, DenseNet201[3] and also explored use of new architecture such as DarkCOVIDNet[1], ConvXNet[2], COVIDNet [5]. A variant to these were use of Monte Carlo drop weights-based network for computing uncertainty in COVID predictions [7].

Interpretability in these works were covered for point estimate networks using GradCAM, GSInquire, Layerwise relevance propagation methods.

There have been limited work on uncertainty estimation for a prediction along with its decomposition and its visualization for interpretability with respect to individual pixels in a Chest X-Ray image, which can be crucial for wider adoption in healthcare. Hence, we develop a visualization framework that captures change in uncertainty with respect to input pixels for a prediction, where the uncertainty was estimated using Bayesian Convolutional Neural Network. We further indicate whether the uncertainty was due to the model (epistemic) or lack of features on the training data not sufficient for classification (aleatoric). Such pixel-level annotation for uncertainty can help healthcare practitioners in understanding the model predictions better, resulting in better trust in the computer-aided diagnosis process.

## 2. PROPOSED FRAMEWORK

There are 3 major components used in our approach:
1. Bayesian Neural Network Architecture [10]
2. Computing Predictive Difference [8]
3. Computing Uncertainty and its components [8][9]

We implement a Bayesian Neural Network [10] that learns the posterior weight distribution p(w|D) given the training data through Dropout Variational Inference, where w aggregates weights over all the L layers in a network and D is the training dataset with N inputs including samples and labels. Similar to [8], we use a more generic alpha divergence than KL divergence to reduce the distance/approximation gap between the true posterior weight distribution and the dropout approximate variational distribution. Since KL divergence underestimates uncertainty due to fitting to local mode of the posterior, we use Alpha divergence that provides better uncertainty estimates relaxing this constraint of KL divergence. In order to determine which input pixels affect the uncertainty of the Bayesian Model, it is important to compute the change in predictive uncertainty, epistemic uncertainty and aleatoric

uncertainty related to a pixel. We use a framework discussed in [8] which is based on the predictive difference analysis [9] to visualize the interpretability through saliency maps visualization. The approach estimates the relevance of each feature in the test input to the prediction by comparing the model output when the feature is known, to the model output when it is not known. The extension of this is applied to estimate the relevance of each feature in test input with respect to the prediction in terms of change in uncertainty. In order to find the relevance of features, a kxk patch of pixels are perturbed to find the change in prediction and uncertainty which unlike gradient based methods yields smoother visualizations.

## 3. EXPERIMENTAL SETUP

Since there is no single benchmark dataset available for COVID Chest X-Ray images, the data was compiled from various open and public repository sources [11][12][13] and segregated to 4 classes for a multiclass classification : COVID-19, Viral Pneumonia, Bacterial Pneumonia and Normal, with a total of 319, 1493, 2772, and 1582 images for each class respectively comprising a mix Anterior to Posterior(AP) and Posterior to Anterior(PA) X-Ray images. Train-test split of 80-20 was done for each class separately due to imbalanced nature of dataset and limited availability of COVID Dataset.

All images were resized to 224x224x3 since each image source had different image sizes. The image pixel values were rescaled to [0,1].

In terms of parameters, 10 stochastic forward passes were performed during training. We used Adam optimizer with a learning rate of 1e-4. Also we choose alpha of 0.5 over 0 or 1 since it proved to yield better predictions and uncertainty estimates [8]. A stopping criterion based on validation loss was set to perform early stopping with patience of 10 epochs, with a total use of 30 epochs. 100 stochastic forward passes are performed during the test phase. An 8X8 feature patch size to be marginalized was chosen with a sliding window with a step size of 10 in both directions to generate smoother visualization. Application of oversampling/data augmentation during initial trials did not yield significant improvement in accuracy, hence it was not considered.

The network resulted in an 80% test accuracy. It is important to note that although the accuracy of similar works [2][3][4] with much more complex networks are around 87%-90%, accuracy was not the focus of our work. The Bayesian network takes considerable amount of computational resources to train. Due to lack of time and limited resources, we stopped hyperparameter tuning with an 80% test accuracy, which we believe is reasonable enough to generate the pixel-level saliency maps, which is the focus of this work.

## 4. RESULTS AND DISCUSSION

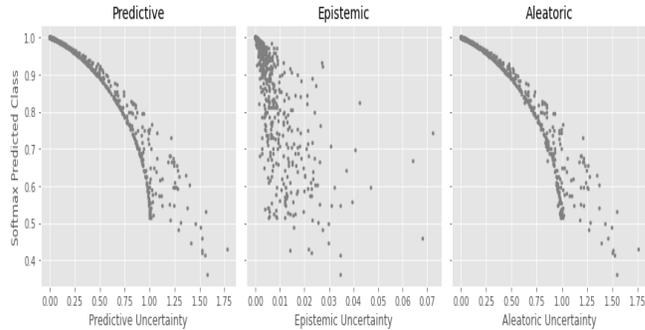

**Figure 1**: Softmax vs Uncertainty for predicted class

Figure 1 shows the relationship between the predictive softmax (max softmax obtained by averaging the softmax values when the input is passed through each weight sample) and the predictive uncertainty, epistemic uncertainty and aleatoric uncertainty for the test samples. It is observed that the predictive uncertainty is higher for lower softmax values and lower for higher softmax values. Decomposing predictive uncertainty into its components (aleatoric and epistemic uncertainty) shows it is mostly influenced by aleatoric uncertainty. This means that the model predictions did not have enough information along the pixels to predict the output value for an input with a fixed weight setting. Also, the softmax for majority of the samples are observed to be above 50%. Epistemic uncertainty for most cases has less influence on the predictions than aleatoric uncertainty. Whereas aleatoric uncertainty is spread more around lower softmax predictions.

Although this is an overall analysis, it is important to analyze each case to understand which pixels when perturbed before passing to the model increased the uncertainty or certainty in the model predictions.

### 4.1. Visualizations for predictive uncertainty and predictive difference:

To demonstrate the usefulness of the visualization tool, we provide explanations through two test image examples, one was predicted correctly, the other one incorrectly. The visualizations generated below for local explanations consists of images that indicate regions of increased uncertainty in red and decreased uncertainty in blue for predictive (indicated in the image as *Diff predictive*), epistemic (indicated in the image as *Diff epistemic*), and aleatoric uncertainties (indicated in the image as *Diff aleatoric*). The scale reflects the change in uncertainty values. In addition to the uncertainty, the predictive difference (indicated in the image as *Diff pred*) of the predicted class is also generated to understand the relationship between predictive difference and predictive uncertainty for each case. Increase in predictive difference/evidence of pixels towards a class is shown in red and decrease in predictive difference/ evidence of pixels against a class is shown in blue. The image content is distributed as Original Image, Change in Uncertainty, and Overlay in order. In the Figures 2 and 3, the right lung is on the left side of the image and left lung on the right side of the image.

### 4.1.1. Example of Correct prediction

Figure 2 shows visualizations generated for a correct COVID prediction. We observe opacities in a continuous fashion on the edges of the right lung. Based on the scale and the absolute value, it could be seen that aleatoric uncertainty is higher than epistemic uncertainty. The model uses mostly the right lung and a smaller area in the left lung to make its predictions for a given class. The same pixels that have contributed to a reduction in aleatoric uncertainty have contributed to reduction in epistemic uncertainty which are along the edges of the right lung and in the middle lobe of the left lung. This shows that the same pixels that provide information for the classification also make the test input similar to the training input distribution. We also notice that in the middle lobe of the right lung, the same pixels contribute to increase in  epistemic and aleatoric uncertainty, showing that the model did not have enough features around the area to make a decision. For COVID case this input is different from the training input with the current combination of features.

When comparing with predictive difference, it can be seen that the pixels that provide evidence for a class reduce predictive uncertainty and pixels that provide evidence against the class increase predictive uncertainty. In this case, the model has correctly picked the regions of difference in densities in the image relevant to ground glass opacities in the right lung. However, the left lung opacities have not been taken into consideration, which requires further pondering into analyzing why the model missed those features and help in model correction and better training samples collection.

### 4.1.2. Example of Incorrect Prediction:

Figure 3 shows visualizations for an incorrect COVID prediction. Based on the scale and the absolute value, it can be seen that aleatoric uncertainty is higher than epistemic uncertainty. The model uses mostly the right lung and a smaller area in the left lung to make its predictions for a given class. Even though the prediction is erroneous, we observe a high softmax value. This particular false positive occurs in some of the sample images because in the training data, some COVID-labeled images were from day 0-2 of COVID. However, in days 0-2 typically the patients can test positive through a swab test, but the anomalies will not show up in a lung X-ray. Hence the model has learnt erroneous features from these noisy training data, where apparently normal lung images were labeled as COVID. Based on such training, for this particular test image, the model has mistaken the change in density due to bone reflections in the right lung middle lobe as opacity and the bronchial vessels as infiltration. Also, the model has picked on the ribs and the background.

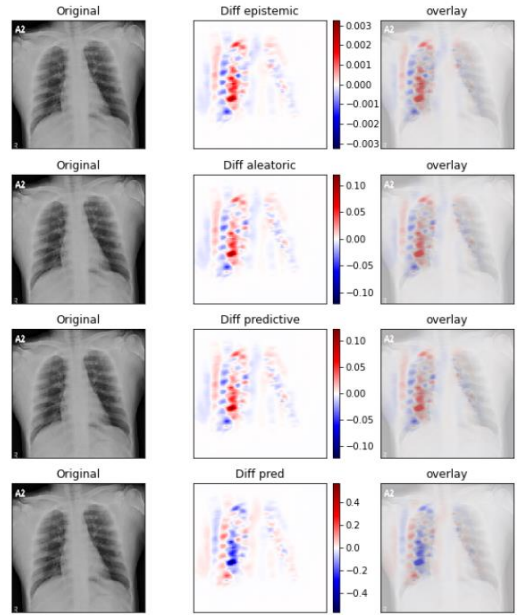

**Figure 2**: Change in epistemic, aleatoric, predictive uncertainty and predictive difference. Ground Truth: COVID, Prediction: COVID, Aleatoric Uncertainty: 0.5657

The pixels adhering to these in the right lung are pixels which are in common reducing epistemic and aleatoric uncertainty, meaning that the same pixels that provide information for the classification also make the test input similar to the training input distribution. We also see that the same pixels contribute for increase in epistemic and aleatoric uncertainty, indicating the testing input has been different with respect to features than the training data that the model is not able to use them for classification. When comparing with predictive difference, it shows that the pixels that provide evidence for a class reduce predictive uncertainty and pixels that provide evidence against the class increase predictive uncertainty.

## 5.CONCLUSION

A visualization framework that demonstrates interpretability based on contribution of pixels in a Chest X-Ray image to uncertainty for COVID 19 diagnosis with local explanations (providing explanation for specific test instance) has been presented, which can help making the computer-aided diagnosis process more trustworthy. The uncertainty estimates are obtained using a Bayesian Convolutional Neural Network and alpha divergence and decomposed to epistemic and aleatoric uncertainty. Aleatoric uncertainty helps identify whether there has been sufficient training data the model has been exposed to, so that it has enough features for a single weight setting to make a classification decision. Epistemic uncertainty demonstrates if there is significant changes in the training and testing data

distribution that can help the radiologist

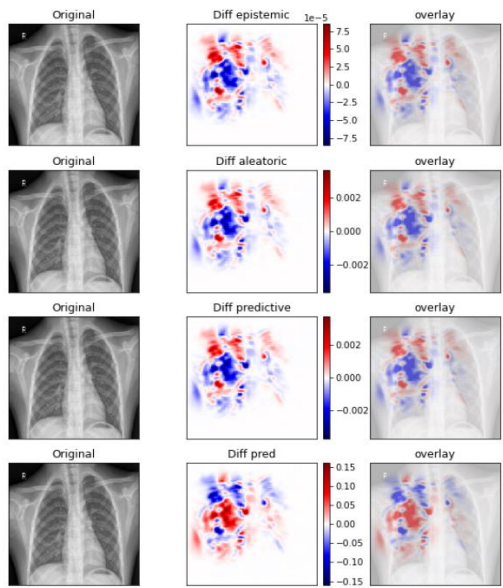

**Figure 3**: Change in epistemic, aleatoric, predictive uncertainty and predictive difference. Ground Truth: Normal, Prediction: COVID, Aleatoric Uncertainty: 0.1284, Epistemic Uncertainty: 0.0015, Softmax: 98.50%

know which type of training data is missing for training and tape appropriate measures to source them.

One of the most prominent limitations in the above work is the lack of a benchmark dataset with enough diversity of COVID cases. Since it is evolving with increasing amount of cases in different parts of the world, diversity in dataset will likely be available in future. The resulting saliency maps depend on the quality of uncertainty estimates from the Bayesian Convolutional Neural Network. This in turn heavily depends on the alpha divergence objective and the approximation gap between the variational distribution chosen to approximate the posterior weight distribution. It is important to make reasonable choices with the distribution and the objective function. An interesting extension would be to explore different variational distributions or divergence objectives and their influence on the uncertainty estimates.